\documentclass{aa}  
\usepackage{graphicx}
\usepackage{txfonts}

\begin{document}

   \title{Spectroscopy of bright quasars: emission lines and internal extinction \thanks{
Based on observations obtained with the 2-m telescope of the Rozhen
National Astronomical Observatory, which is operated by the Institute
of Astronomy, Bulgarian Academy of Sciences, and with the 1.3-m telescope
of the Skinakas Observatory, Crete, Greece; Skinakas Observatory is a
collaborative project of the University of Crete, the Foundation
for Research and Technology -- Hellas, and the Max-Planck-Institut
f\"ur Extraterrestrische Physik.
}}

   \author{R. Bachev \inst{1} 
          \and A. Strigachev \inst{1,2} 
          \and E. Semkov \inst{1} 
          \and B. Mihov \inst{1}}

   \offprints{R. Bachev}

   \institute{Institute of Astronomy, Bulgarian Academy of Sciences, 72 Tsarigradsko Chausse Blvd., 
               1784 Sofia, Bulgaria\\
              \email{bachevr@astro.bas.bg}
   \and Royal Observatory of Belgium, Av. Circulaire 3, B-1180 Brussels, Belgium \\
                  \email{Anton.Strigachev@oma.be}}
   \date{Received ...; accepted ...}

\titlerunning{Spectroscopy of bright quasars}
\authorrunning{R.Bachev et al.}

  \abstract
   {}
   {The main purpose of this work is to improve the existing knowledge about the most 
powerful engines in the Universe -- quasars. Although a lot is already known, we still 
have only a vague idea how these engines work exactly, why they behave as they do, and 
what the relation is between their evolution and the evolution of their harboring galaxy.}
   {Methods we used are based on optical spectroscopy of visually bright quasars, many of
which have recently been discovered as X-ray sources, but eventually missed in color-selected 
surveys. The spectra typically cover the 4200--7000 \AA\AA\ region, allowing 
measurements of the characteristics of the hydrogen lines, the FeII contribution, and 
other lines of interest.}
   {We present accurate redshift estimates and Seyfert type classification of the 
objects. We also show that the contribution of the host galaxy to the optical continuum 
is non-negligible in many cases, as is the intrinsic AGN absorption. Consequences 
of not correcting for those factors when estimating different quasar parameters are 
discussed. We also find some evidence of a non-unity slope in the relation between the 
internal extinction based on the Balmer decrement and the one on the optical continuum slope, 
implying, if further confirmed, the intriguing possibility that some absorbing material 
might actually be located $between$ the continuum source and the broad-line region.}
   {}


\maketitle

\section{Introduction}
The generally accepted paradigm about the nature of quasars (or active galactic nuclei; AGN) 
invokes a supermassive black hole, surrounded by an accretion disk, where most of the emitted 
energy is generated. In the vicinities of this disk, there must be a region, producing the 
broad emission lines (BLR), traditionally presented as a system of clouds, moving in the 
gravitational field of the black hole. Farther out from the center should be a region 
producing narrow lines (NLR), which may be the only visible optical lines in the spectra, 
if the central region is hidden from the observer behind a thick torus, as the unification 
models explain the AGN type I/II differences (Antonucci, 1993, for a review). 
Studying the narrow lines is important, since they are fairly sensitive to the shape 
of the ionizing continuum, and especially to the unobserved EUV range, which, if restored 
properly, may help to test the general paradigm or to model the type of accretion disk 
(e.g. standard thin disk, ADAF, etc.).
One of the least clear points in this picture so far appears to be the exact geometry 
(e.g. flat, spherical, conical), structure (system of clouds, irradiated accretion disk, 
etc.), and kinematics (inflow, outflow, Keplerian rotation) of the BLR matter. The BLR should 
play an important role in feeding the quasar; therefore, knowledge of its nature may 
reveal why some galactic centers are active and others are not, in case a supermassive 
black hole is likely to be present in both types (Magorrian et al.  1998).

The best way to probe the broad-line region might be through its optical emission-line 
spectrum. Although a single spectrum is certainly not sufficient to answer all the questions, 
studying the line profiles, ratios, and other properties in many objects seems to be one 
way of sheding some light on the problem. A promising new approach to systematizing quasar 
properties and differentiating between quasars is based on principle component analysis 
(PCA), where the quasar diversity can be expressed as a linear combination of a few 
eigenvectors of the correlation matrix, constructed from different measured quantities 
of a sample of quasars. Many of these measures are based on the optical spectra 
(Boroson \& Green 1992; Sulentic et al. 2000), such as the width of the broad lines, 
equivalent width ratios. It is proposed that the main eigenvector (EV1), i.e. the
main driver of the quasar diversity, is the accretion rate ratio (Marziani et al. 2001),
one of the defining characteristics of the AGN.
On the other hand, another important characteristic of the central engine, the black 
hole mass, is often calibrated through empirical relations, like the ``width of the broad 
lines, FWHM -- the monochromatic continuum luminosity, $\lambda L_{\lambda5100}\AA$'', 
(Kaspi et al.  2005), ``the black hole mass -- bulge luminosity or bulge velocity 
dispersion'' (Magorrian et al.  1998; Ferrarese \& Merritt, 2000), etc., since the 
only direct method known, the reverberation mapping (see Peterson \& Horne, 2004, for a review), 
is observationally very expensive. It is clear that a large sample of quasars with relatively 
good spectra is needed for better understanding the importance of EV1 correlations for the quasar 
physics and to better calibrate different quasar empirical relations.

Nowadays we see that many bright broad-line quasars have been missed in color-selected samples 
(e.g. Palomar--Green, Schmidt \& Green, 1983), mainly because of their reddened colors, 
due to significant host galaxy contribution and/or intrinsic absorption. 
Being mostly radio-quiet, many of those objects have only recently been discovered and 
identified as quasars through X-ray surveys. Studying such objects in detail is 
especially important for preventing introduction of different biases into our knowledge 
about the overall AGN population, its properties, and evolution. Furthermore, these 
partially obscured objects (reddened S1, S1.8, S1.9 types, etc.) may provide an 
important link between the ``pure'' type I AGN and the obscured population (type II / 
Seyfert 2), especially taking into account the building evidence that the latter may 
comprise a significant part of the entire AGN population (Zakamska et al.  2003).
This paper should be considered as a part of a larger continuing effort to obtain spectra 
of relatively good quality for most of the nearby, visually bright quasars, including 
many discovered in recent years. In a subsequent paper, we will concentrate on the 
connection between the quasar host galaxies and different AGN characteristics.

The paper is organized as follows. In the next two chapters we describe the observations 
and present the main results, including individual notes on the objects. Discussion 
is presented in Sect. 4 and the summary, in Sect. 5.

\begin{table*}
\caption{Log of observations.}
\label{table:1}      
\centering          
\begin{tabular}{lcccccccccc}
\noalign{\smallskip}
\hline\hline
Object & RA J2000 & Dec J2000 & z & $\sigma$(z) &  A$_{\rm V}$ & Date & Instr. & Exp. & S/N\\
\hline

2MASXJ005050+3536& 00 50 51 & +35 36 43 & 0.0585 & 0.0004 & 0.14 & 08.09.2004 & Rz 2.0 &   2700 &  36\\
MCG+08.17.060    & 09 13 46 & +47 42 00 & 0.0524 & 0.0005 & 0.05 & 19.03.2004 & Rz 2.0 & 2x2400 &  10\\
RXS J11401+4115  & 11 40 03 & +41 15 05 & 0.0717 & 0.0002 & 0.06 & 22.06.2005 & Sk 1.3 & 2x1200 &  12\\
PG 1211+143      & 12 14 18 & +14 03 13 & 0.0815 & 0.0003 & 0.11 & 23.06.2005 & Sk 1.3 & 2x1200 &  27\\
RXS J12308+0115  & 12 30 50 & +01 15 21 & 0.1183 & 0.0005 & 0.06 & 24.06.2005 & Sk 1.3 &   1200 &  25\\
RXS J16312+0955  & 16 31 16 & +09 55 58 & 0.0917 & 0.0005 & 0.18 & 17.08.2004 & Sk 1.3 & 2x2700 &  35\\
RXS J17233+3630  & 17 23 25 & +36 30 25 & 0.0400 & 0.0003 & 0.16 & 29.06.2005 & Rz 2.0 &   1800 &  25\\
IRAS 18423+2201  & 18 44 30 & +22 04 28 & 0.0464 & 0.0004 & 0.69 & 11.08.2005 & Sk 1.3 & 2x2400 &  15\\
NPM1G+27.0587    & 18 53 04 & +27 50 28 & 0.0620 & 0.0004 & 0.49 & 10.08.2005 & Sk 1.3 & 2x1800 &  31\\
                 &          &           &        &        &      & 04.05.2005 & Rz 2.0 &   1800 &  29\\
RXS J20440+2833  & 20 44 04 & +28 33 09 & 0.0492 & 0.0005 & 1.03 & 28.08.2005 & Sk 1.3 &   2400 &  38\\
                 &          &           &        &        &      & 30.06.2005 & Rz 2.0 &   1800 &  33\\
NPM1G$-$05.0589  & 21 03 38 &$-$04 55 40& 0.0575 & 0.0002 & 0.27 & 18.08.2004 & Sk 1.3 & 2x2700 &  31\\
RXS J21240$-$0021& 21 24 02 &$-$00 21 58& 0.0617 & 0.0003 & 0.15 & 17.08.2004 & Sk 1.3 & 2x2700 &  36\\
NPM1G+24.0470    & 21 39 41 & +24 24 18 & 0.0390 & 0.0005 & 0.23 & 10.08.2005 & Sk 1.3 &   2400 &  26\\
RXS J21592+0952  & 21 59 12 & +09 52 42 & 0.1003 & 0.0002 & 0.21 & 19.08.2004 & Sk 1.3 & 2x2700 &  39\\
RXS J22027$-$1304& 22 02 45 &$-$13 04 53& 0.0391 & 0.0003 & 0.14 & 28.08.2005 & Sk 1.3 & 2x2400 &  22\\
RXS J22160+1107  & 22 16 04 & +11 07 26 & 0.0514 & 0.0005 & 0.21 & 10.08.2005 & Sk 1.3 &   2400 &  16\\
RXS J22287+3335  & 22 28 46 & +33 35 08 & 0.0906 & 0.0004 & 0.29 & 11.08.2005 & Sk 1.3 & 2x2400 &  35\\
NPM1G$-$04.0637  & 22 53 11 &$-$04 08 49& 0.0257 & 0.0002 & 0.14 & 11.08.2005 & Sk 1.3 & 2x2400 &  24\\

\hline\hline
\end{tabular}
\end{table*}

\section{Observations and reductions}

\subsection{Objects}
We did not apply any formal criteria to select the objects. Our intention was mostly to 
observe Type I objects from the V\'{e}ron-Cetty \& V\'{e}ron catalog (2003; hereafter VCV03) 
with no previously published spectra of good quality and resolution. The objects were 
bright enough to be accessible with 1.5--2-m class telescopes (Sect. 2.2), i.e. -- 
14--15-th magnitude. Most of these objects were discovered initially as X-ray 
sources (RASS catalog) and identified as quasars / Seyfert galaxies only recently. 
One can naturally expect that these objects will appear to be predominantly 
radio quiet, as in our case. A few quasars, like PG 1211+143, are well-studied objects; 
however, they have been observed to be able to better assess our measurement errors and 
the quality of spectral calibration and to look for possible variations. The most common 
names of the objects and their J2000 coordinates, as given by VCV03 are shown in Table 1 
(cols. 1 -- 2). The next columns in the table contain our measured redshifts with the errors 
(Sect. 2.3 for details) and the Galactic absorption.

\subsection{Observational data}
The last four columns in Table 1 show the observation date, the instrument (Rz 2.0 is the Rozhen telescope, 
Sk 1.3 is the Skinakas telescope, see the text below), the exposure time in seconds and the signal-to-noise 
ratio.
The spectra of the objects were obtained with two instruments -- the 2-m RCC telescope of the 
Rozhen National Observatory, Bulgaria, and the 1.3-m RC telescope of Skinakas Observatory, 
Crete. The Rozhen telescope is equipped with a Photometrics AT200 CCD camera (1024$\times$1024 
pixels) and the focal reducer FoReRo attached to the RC focus of the telescope. 
A grating prism (300 lines/mm) and 130 $\mu$m slit were used. The Skinakas telescope is 
equipped with a 2000$\times$800 ISA 612 CCD attached to a focal reducer. 
A grating prism of 1300 lines/mm and a slit of 160 $\mu$m were used there. 
 
A standard HEAR lamp or nightsky lines were used for the wavelength calibration, allowing a wavelength 
accuracy typically better than 2--3 \AA\ (Table 1).  
Suitably located spectroscopic standards were used for the flux calibration. 
The typical signal-to-noise ratio of the spectra is 25--30 and the spectral resolution --
5--10 \AA\ (Table 1).

\subsection{Reductions}

We used standard IRAF routines to perform wavelength and flux calibrations and to extract 
1-D spectra of the objects. After that, each spectrum has been corrected for the atmospheric 
A and B-band absorption, which in a few cases, unfortunately, appeared very close to strong 
emission lines of interest. The results from this correction are rather satisfactory 
for the majority of the objects. All spectra were also corrected for the Galactic 
absorption (Schlegel et al. 1998), which although typically rather small 
($A_{\rm V} \simeq 0.1 - 0.2$ mag.), turned out to be significant for some objects 
-- in order of a magnitude (Table 1). 
The [OIII]$\lambda5007$ line was mostly used to calculate the redshift and to perform 
the redshift correction, since it was often the most prominent narrow line in the spectrum. 
In a few spectra, where [OIII] was weak (and likely to be blue-shifted, e.g. 
Zamanov et al.  2002), we used the narrow top of H$\beta$ and/or H$\alpha$. 
The redshifts are listed in Table 1. We also show typical errors, which are based on 
the quality of the spectrum and on how well redshifts of different lines 
agree in between. Our typical redshift error is $\simeq0.0005$ or less, and the standard 
deviation between our redshifts and those from VCV03 is $\simeq0.001$.

\section{Results}

\subsection{Host galaxy and FeII contributions}
The redshift-corrected spectra of the objects included in this study are shown in Figs. 1 
and 2. To be able to measure the line profiles and the intrinsic power-law 
continuum accurately, two more corrections had to be made. 
The first one is the subtraction of the 
host galaxy (starlight) contribution to the continuum. The host galaxy signatures are 
clearly seen in many objects that finally turned out to be S1.8/S1.9 type galaxies 
and, to some extent, in S1 types as well. 
One can approximately estimate the host galaxy contribution by the 
strength of the stellar absorption lines, such as the CN g-band (around 4300\AA), MgB triplet 
(around 5172\AA), Ca+Fe (arond 5269\AA), and NaD lines (5892\AA). We subtracted a 
scaled S0 template from each spectrum, made publicly available by Kinney et al. (1996); 
see Hao et al. (2005) and Strateva et al. (2005) for an alternative approach. 
The S0 galaxy was arbitrary chosen at first, because we have little knowledge about the exact 
host galaxy type in most cases, and second, because many spectra are rather noisy to be able 
to precisely distinguish the templates, and finally, because the differences 
between the templates are actually not that large. The scaling of the template was chosen 
in such a way that no signs of absorption lines are seen in the resulting spectra. 
To judge the strength of the host galaxy contribution, the 
characteristic shape of spectra in the 4200--4600 \AA\AA\ range, where the host 
galaxy continuum changes significantly, is also helpful.
Of course, such an approach is prone to subjectivity, which in addition to possible 
systematic errors (mismatched galaxy type, possible extra starburst component, 
possible problems with the flux calibration of the spectra, noise, etc.), does not 
allow us to make any strong statements about the exact value of the host galaxy 
contribution. It can, however, give some good idea about how much this contribution 
can be, and indeed, our data suggest that it is significant in many cases 
(Fig. 1, 2; Sect. 3, 4). For a few objects, the host galaxy subtraction helped to reveal 
broad wings of the H$\alpha$ and/or H$\beta$ lines, clearly seen only on the remaining 
flat continuum after the subtraction, which is very important for the correct 
classification of the objects.

\begin{figure*}
\centering
\resizebox{\hsize}{!}{\includegraphics{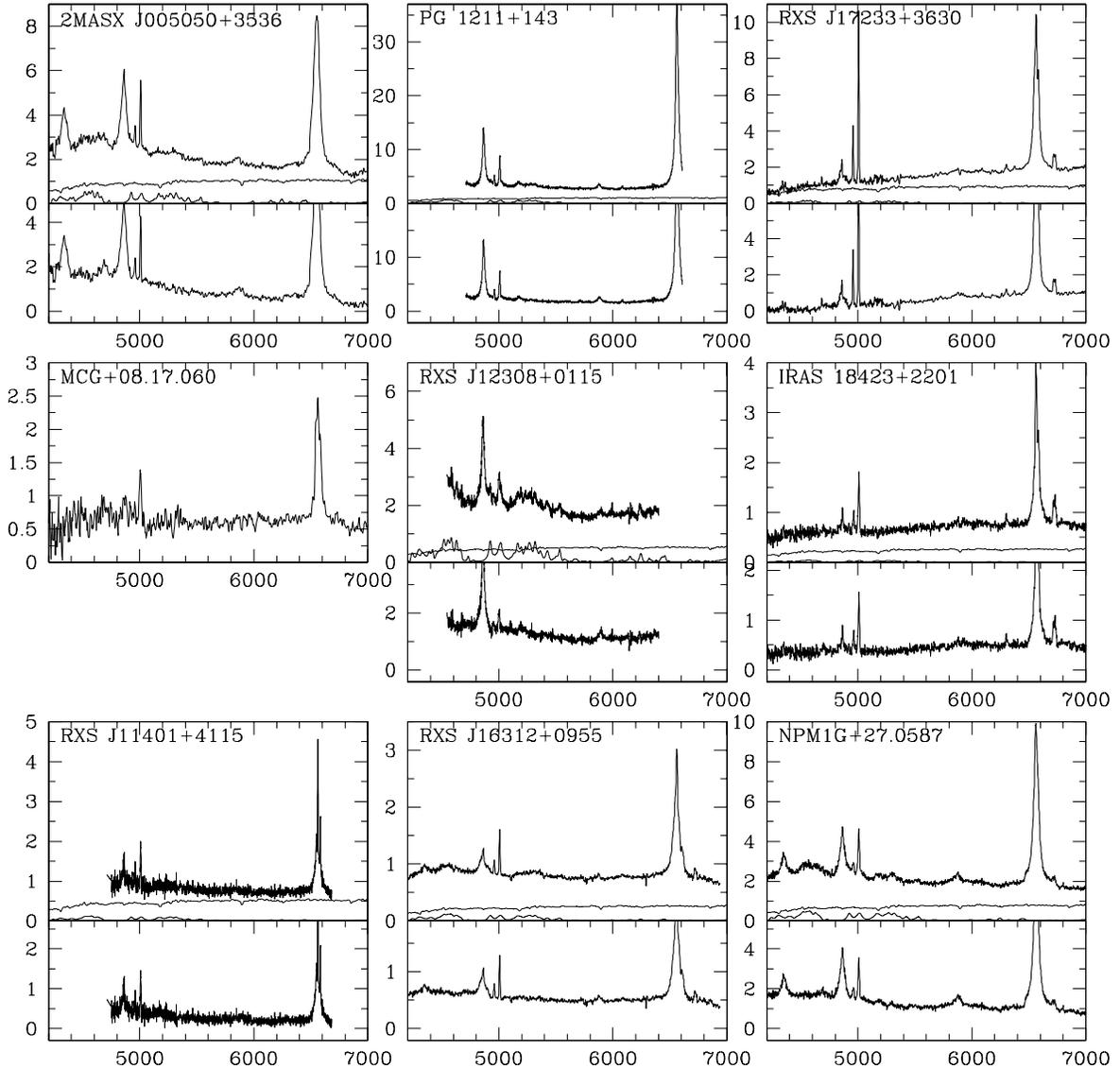}}
\caption{Optical spectra of all objects from our sample. All spectra are corrected 
for atmospheric absorption bands, Galactic absorption and are deredshifted. The upper part of 
each box displays the original spectrum (thick line) and the FeII and starlight templates to be 
subtracted from it (thin lines). The lower part shows the resulting intrinsic AGN spectrum.
All fluxes are in units of $10^{-15}$ ergs$^{-1}$cm$^{-2}$\AA$^{-1}$.}
\label{Figure:1}
\end{figure*}

\begin{figure*}
\centering
\resizebox{\hsize}{!}{\includegraphics{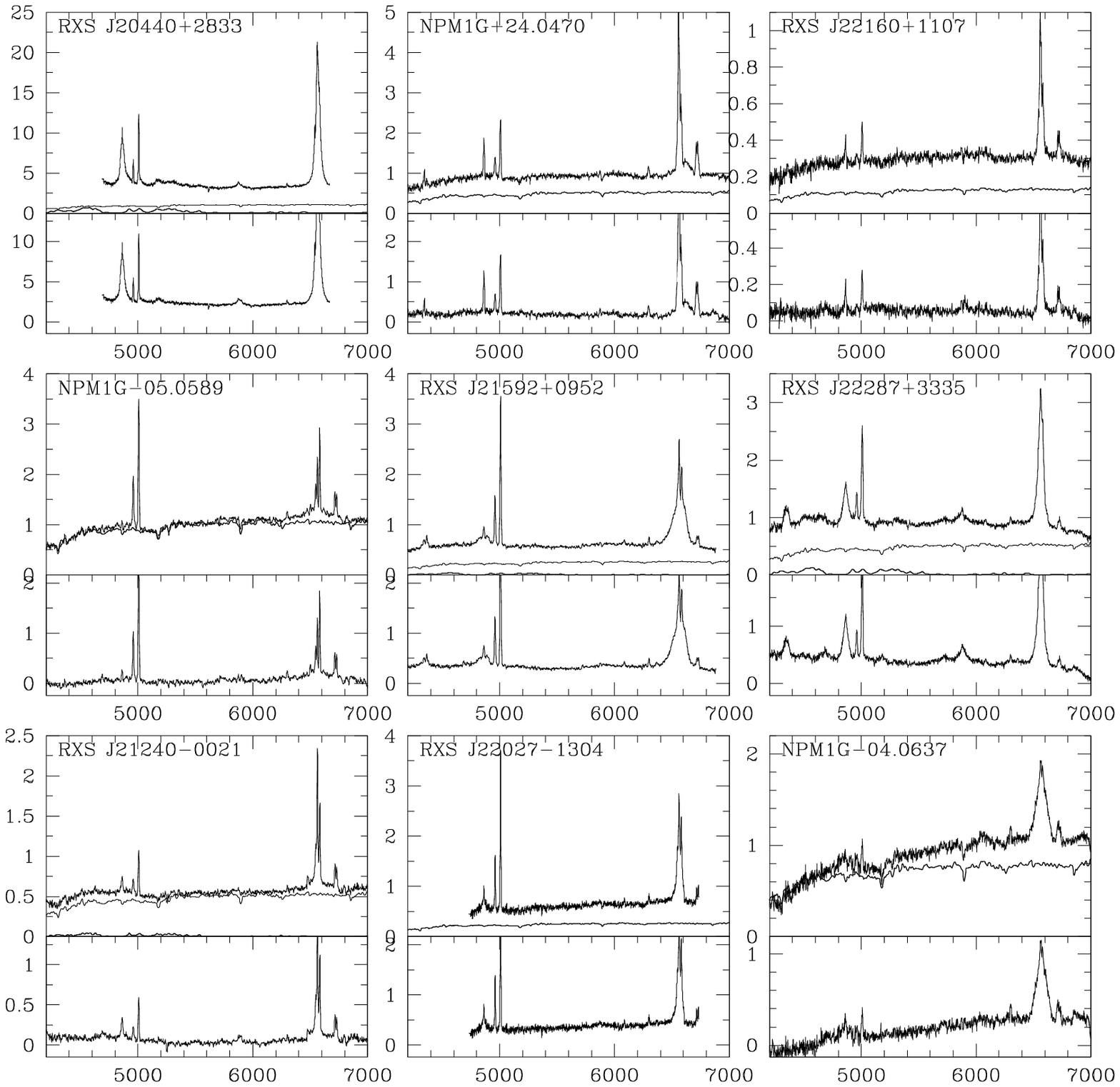}}
\caption{See Fig.1.}
\label{Figure:2}
\end{figure*}

Another ``contaminating'' contribution is from the FeII lines. Of course, being 
intrinsic to the BLR, these lines cannot be considered as a contamination, but their
subtraction is necessary for two reasons. First, it allows better measurements of 
lines like HeII$\lambda4686$, [OIII]$\lambda4959,5007$, and to some extent, H$\beta$. 
After the FeII template subtraction, a broad feature, possibly a blend of 
[FeVII]$\lambda5160$, [FeVI]$\lambda5177$, and [NI]$\lambda5200$ is 
clearly seen in many objects at a level roughly about 20\% from the FeII 
flux (Table 3). This feature is otherwise completely hidden in the template. 
It may, indeed, be partially due to a poor host galaxy subtraction. Second, the EV1 
correlations reveal the significance of the FeII emission, where 
$R_{\rm Fe}$=F(FeII)/F(H$\beta_{\rm broad}$) is one of the most important parameters 
used there. Therefore, the FeII contribution has to be estimated as precisely as possible.
We used a broadened and scaled FeII template of I Zw 1 as is usually done 
(see for instance Marziani et al.  2003, for more details on the FeII subtraction). 
Again, no formal criterion was used to estimate the iron strength, and it has to 
be considered as a best guess with an error typically of about 20\% or higher for the 
noisier spectra. One can judge how successful the FeII subtraction was by the 
flatness of the continuum around H$\beta$ (Figs. 1, 2, the lower panel of each box). 
The FeII emission flux ratio ($R_{\rm Fe}$) is given in Table 2. 
The FeII flux is obtained by integrating the template between 4434 and 4684\AA\AA, as 
in Boroson \& Green (1992). This is only $\sim$25\% of the total 
FeII emission between 4000 and 7000\AA\AA, so for many objects, iron has to be 
considered as producing the most prominent line features in optics, not hydrogen. 
The order of the host galaxy and the FeII subtractions was not the same for all 
the objects; we started with the template, whose contribution seemed more significant. 
Figures 1 and 2 show all the spectra. The upper panel of each box displays the original 
spectrum and the templates to be subtracted from it. The lower panel shows the spectrum 
after the subtractions, used further for the emission line and continuum measurements 
(Sect. 3.2). We do not try to fit a power law continuum to the 
resulting quasar spectrum (e.g. Dietrich et al.  2005), since there is no guarantee
that a single power-law describes the entire AGN optical continuum (although we do
measure a local slope, Sect. 3.5).

\subsection{Emission-line measurements}
Fluxes (in units of $10^{-15}$ ergs$^{-1}$cm$^{-2}$\AA$^{-1}$) of various lines as measured from the 
spectra are shown in Tables 2 and 3. All widths are in kms$^{-1}$, $\lambda L_{\rm \lambda}$ is in 
ergs$^{-1}$, and the black hole masses (Table 2, the last column, see also below) are in Solar units.
In Tables 2 and 3 question marks indicate insecure line measurements or lines, too noisy to measure 
accurately. When no line is seen in the spectrum, the corresponding box is empty. The typical errors 
of the optical slopes and the central masses (Table 2, the last two columns) are estimated to be $\pm 0.5$.

The narrow lines (Table 3) are measured after fitting with a Gaussian. 
[FeVI]$\lambda$5177 is probably blended with [FeVII]$\lambda$5160 and [NI]$\lambda$5200.
The typical errors of the narro-line measurements are $\sim 30\%$.
No narrow line widths are shown, since the widths are often determined by the 
instrumental width (or we cannot verify the opposite). We also do not show the 
equivalent widths (EW), due to the uncertainty of the continuum level, often 
dominated by the host galaxy (see above), in which case EW hardly has any meaning. 
Where present, the broad lines (H$\alpha$ and H$\beta$) were measured 
after an attempt to subtract the narrow component on top. We succeeded in isolating 
a narrow component in practically all H$\beta$ profiles, where in some cases 
inflection was apparent, while in other -- we subtracted our ``best guess'' for the 
narrow component, which was supposed to have the same width as the [OIII] lines. 
The errors in the latter case can be, of course, significant (as much as 50\% -- 
much more than the typical error of [OIII], which is about 10\%), at least 
because a narrow top does not necessarily mean an NLR emission. A similar approach 
was used for H$\alpha$ lines; however there, the blended [NII]$\lambda6548,6583$\AA\ 
lines could not be separated from the H$\alpha$ profile in many cases, especially 
when the H$\alpha$ profile itself was broad enough (see Sect. 3.5 for 
details on the individual spectra). In those cases, only a broad component was 
measured, and one should be aware about some possible overestimation of 
both -- the flux and the FWHM of H$\alpha$ (actually FWHM can also be underestimated 
if the narrow component of H$\alpha$ is much stronger than the [NII] lines). 
Where possible, the narrow component of H$\alpha$ and NII lines were 
deblended and they, together with the broad component, were measured separately. 
The parameters of the broad components 
(FWHM, flux) were measured after fitting with Gaussian or Lorentzian functions. 
The choice was made upon the goodness of the fit and is indicated in Table 2. 
For most of the objects, one of these two possibilities worked quite well (Sect. 4.1).
The measurement errors of the FWHM and of the flux of the broad components are typically 
10--20\%, estimated after different fits and continuum levels were attempted.

\begin{table*}
\caption{Broad-line measurements.}

\label{table:2}
\begin{tabular}{lccccccccccc}
\hline\hline

Object & Flux & FWHM & fit & R$_{\rm Fe}$ & Flux & FWHM & Flux & Flux  & log$\lambda L_{\lambda}$ & 
$a_{\rm \nu}$ & logM$_{\rm BH}$ \\

 & H$\beta$ & H$\beta$ & H$\beta$ & & H$\alpha$ & H$\alpha$ & HeII4686 & HeI5876 & 5100\AA & &  \\

\hline

2MASXJ005050+35 & 180 & 3690 & G & 0.49 & 460 & 3060 & 39 & 37 & 44.0 & 2.7 & 7.5 \\
MCG+08.17.060   & ? & ? & & ? & 92 & 2740 & ? & ? & 43.6 & ? & ? \\
RXS J11401+4115 & 21 & 3130 & G? & 1.05 & 72 & 2290 & & &  43.7 & 1.4 & 7.2 \\
PG 1211+143 & 415 & 1810 & L & 0.30 & 1400 & 1280 & & 38 & 44.6 & $-$0.5 & 7.3 \\
RXS J12308+0115 & 106 & 2120 & L? & 1.25 & & & 2 & ? & 44.7 & $-$0.6 & 7.5 \\
RXS J16312+0955 & 23 & 4380 & G & 0.96 & 113 & 2290 & 3 & 5 & 44.1 & $-$1.4 & 7.7 \\
RXS J17233+3630 & 55 & 4920 & G & $\leq$0.5 & 370 & 2600 & 15 & 21 & 43.2 & $-$5.8 & 7.2 \\
IRAS 18423+2201 & 14 & 3810 & G & $\leq$0.5 & 90 & ? & 4 & 10? & 43.3 & $-$3.4& 7.1 \\
NPM1G+27.0587 & 113 & 3870 & G & 0.78 & 460 & 2740 & 25 & 75 & 44.1 & 0.4 & 7.6 \\
RXS J20440+2833 & 268 & 2500 & L? & 0.46 & 837 & 1690 & & 44 & 44.2 & $-$1.0 & 7.3 \\
NPM1G-05.0589 & & & & -- & $<$30 & 11450? & 2 & & -- & -- & -- \\
RXS J21240-0021 & 6 & 4320 & G & $<$0.1 & $<$24 & 6850? & 5 & 3? & 42.8 & $-$0.2 & 6.8 \\
NPM1G+24.0470 & & & & -- & 27 & 6220 & 0 & 3 & 42.9 & $-$0.7 & -- \\
RXS J21592+0952 & 20 & 6950 & G & 0.36 & 133 & 7400 & 2 & ? &  43.9 & $-$2.0 & 8.0 \\
RXS J22027-1304 & 15 & 4750 & G & $<$0.1 & 50 & 2560 & & 7 &  43.1 & $-$3.2 & 7.1 \\
RXS J22160+1107 & 3 & 3310 & G & - & 13 & 3650? & 4 & ? & 42.6 & $-$0.4 & 6.4 \\
RXS J22287+3335 & 36 & 3630 & G & 0.61 & 130 & 2610 & 12 & 25 &  43.9 & $-$0.7 & 7.4 \\
NPM1G-04.0637 & 14 & 7250 & G & $<$0.1 & 84 & 4750 & 9? & 0 & 42.2 & $-$7.6 & 6.9 \\

\hline\hline

\end{tabular}
\end{table*}


\begin{figure}
\centering
\resizebox{\hsize}{!}{\includegraphics{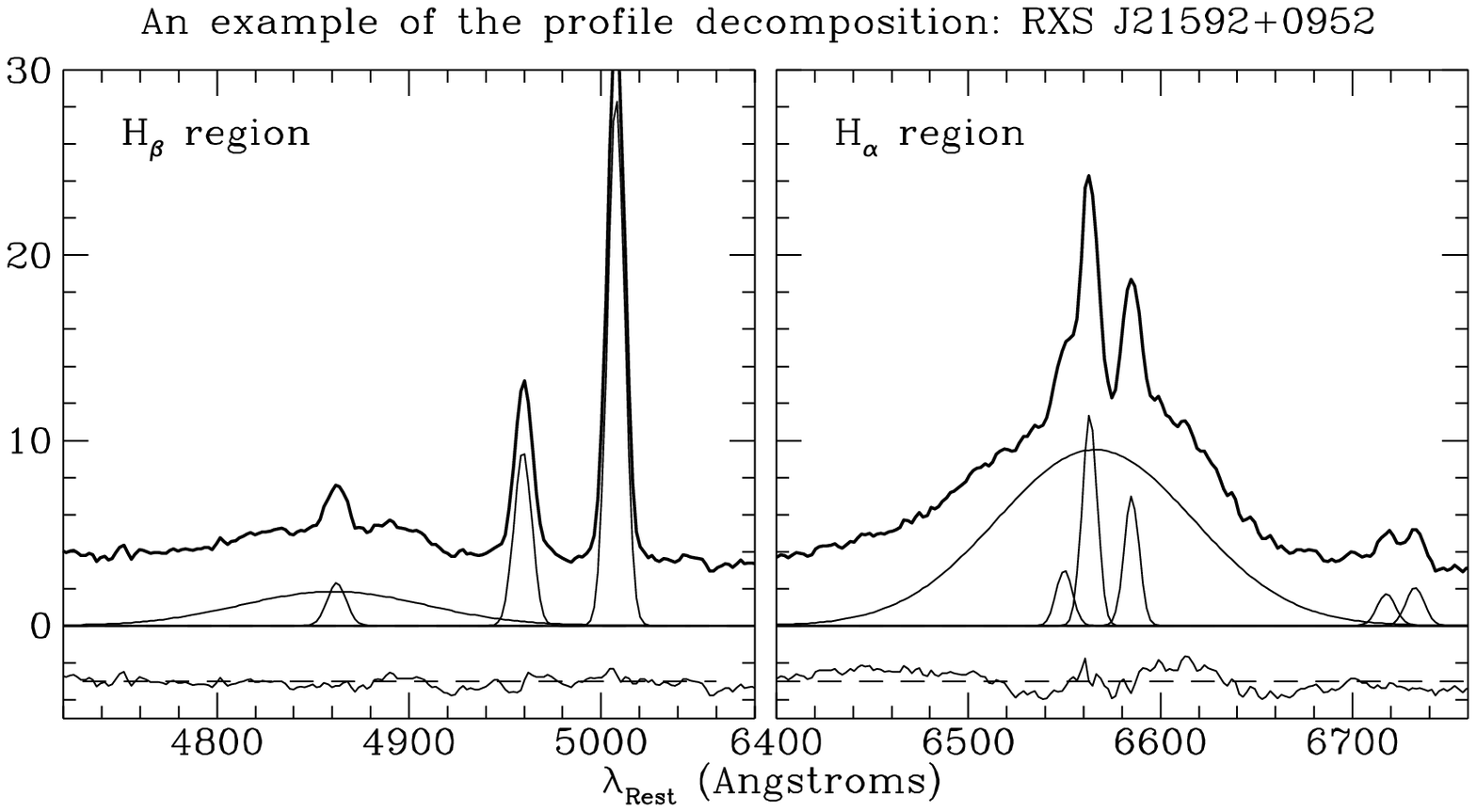}}
\caption{An example of profile decompositions for one of the objects (RXS J21592+0952). The thick line 
(upper panels) represents the original spectrum (after galaxy and FeII subtractions). The thin 
lines are the fitted lines, where only Gaussians are used. The residuals are shown in the 
bottom panels. The left panels are for the H$\beta$ region ([OIII] and H$\beta$ broad and narrow 
components are shown) and the right panels -- the H$\alpha$ region ([SII], [NII] 
and both components of H$\alpha$ are shown). One sees that a single function (Gaussian or Lorentzian) 
does not always model the broad emission perfectly (which is not the purpose of this paper), 
as seen from the residuals for the H$\alpha$ region.
The fluxes are in $10^{-16}$ ergs$^{-1}$cm$^{-2}$\AA$^{-1}$}
\label{Figure:3}
\end{figure}

Figure 3 shows an example of profile decompositions for one of the objects (RXS J21592+0952). 
Our sample seems to confirm that the narrower-line objects (and eventually those with stronger 
FeII emission) tend to be better fitted by a Lorentzian, and vise versa (Sulentic et al.  
2000; V\'{e}ron-Cetty et al.  2001). See however the Sect. 4.1 for comments on this issue.
Table 2 also contains the monochromatic flux at 5100\AA, often used for BLR size 
($R_{\rm BLR}$) and central mass ($M_{\rm BH}$) estimates. 
The value of $\lambda L_{\lambda5100}$ is measured on the resulting spectrum after the 
FeII and starlight corrections and should be considered as a measure of the 
intrinsic quasar optical continuum. Due to the uncertainties of those corrections, 
the uncertainties of $\lambda L_{\lambda5100}$ have to be in the 30--40\% range. 
Knowing the width of the lines and $\lambda L_{\lambda5100}$, we can estimate 
the black hole mass for each S1 object (Table 2), assuming

$\log(M_{\rm BH})= 0.69 \log(\lambda L_{\rm \lambda}) + 2 \log(FWHM(H \beta ))-30.0$, 
obtained following Kaspi et al. (2005). 
No corrections for the intrinsic absorption have been made, and if 
significant, the absorption can lead to underestimating the continuum level, and 
to underestimating of $M_{\rm BH}$ up to a magnitude or even more (Sect. 4).
An alternative calibration (Bentz et al.  2006), taking the host galaxy 
contribution to the monochromatic luminosity into account, leads to a slightly different 
expression for the black hole mass, but does not alter our results significantly.

The last column of Table 3 shows the type of the object (e.g. S1/2, LINER, HII galaxy), 
based on the presence of a broad component or different narrow line (diagnostic) ratios.
Although all the objects were initially classified as S1 types (according to VCV03), 
our spectra clearly indicate that some of them should probably be classified 
as S1.9/S2/LINER types (Table 3). 
To be able to distinguish between the narrow-line 
objects (e.g. S2, LINER and HII galaxies), we employed diagnostic diagrams 
(Veilleux \& Osterbrock, 1987; V\'{e}ron-Cetty et al.  2001), where ratios like 
[OIII]$\lambda5007$/H$\beta_{\rm narrow}$, [SII]$\lambda$6716+6731/H$\alpha_{\rm narrow}$, 
[OI]$\lambda6300$/H$\alpha_{\rm narrow}$, etc. are used to separate objects of different types. 
We also checked the place on these diagrams of some broad wing objects (i.e. S1.8, S1.9) to 
look for possible LINER 1 type objects (i.e. LINERs with broad line wings), whose 
existence has been suggested by Ho et al. (1997a, b). We found, however, no 
clear evidence of such objects in our sample. 
See Sect. 3.5 for classification details.

\begin{table*}
\caption{Narrow-line flux measurements in units of $10^{-15}$ ergs$^{-1}$cm$^{-2}$\AA$^{-1}$.}

\label{table:3}
\begin{tabular}{lcccccccccc}
\hline\hline

Objects & H$\beta$ & [OIII]& [FeVI]       & [FeVII]&[OI] & H$\alpha$ & [NII]& [NII] & [SII]   &   AGN type\\
        & 4861     & 5007  & 5177         & 6087   &6300 & 6563      & 6548 & 6583  & 6716+31 &           \\             
\hline

2MASXJ005050+3536  & 11 & 30 &   &   & 11& & & & 0 & S1 \\
MCG+08.17.060      & ?  & 17 &   &   & 3?& 4 & 3? & 5 & 4 & S1 \\
RXS J11401+4115    & 4.8& 7.8& 5 &   & 2 & 15 & 2.2 & 8.5 & & (LL)S1 \\
PG 1211+143        & 25 & 45 & 23& 74& 0 & & & & & NLS1\\
RXS J12308+0115    & 8? & 9.6&   &   & & & & & & (NL)S1\\
RXS J16312+0955    & 1.6& 8  &   &   & ? & & & & 4 & S1 \\
RXS J17233+3630    & 11 & 94 &3.1&3.5& 8 & 32 & ? & 18 & 20 & S1 \\
IRAS 18423+2201    & 4.3& 15 &   &   & 5.2 & 21? & & & 13 & S1 (L1)\\
NPM1G+27.0587      & 8.2& 27 & 5 &   & 5 & & & & 13 & S1 \\
RXS J20440+2833    & 10 & 70 & 30&   & 9 & & & & & S1\\
NPM1G-05.0589      & 2.6& 30 &   &   & 2 & 10 & 7 & 16 & 9.5 & S1.9? \\
RXS J21240-0021    & 2.5& 6  &   &   & 1? & 10? & 4.6 & 9.1 & 6.5 & S1 (LLS1, L1)\\
NPM1G+24.0470      & 12 & 19 &   &   & 3.5 & 60 & 10? & 29 & 19.8 & S1.9? (L1, HII)\\
RXS J21592+0952    & 2.8& 33 &   & 2 & 2.6 & 11 & 3 & 7 & 4.6 & S1 \\
RXS J22027-1304    & 2.2& 27 &   & 2 & 2.3 & 9 & 5? & 9 & 5 & S1.8\\
RXS J22160+1107    & 1.1& 2.4&   &   & & 6? & 2.7 & 5.8 & 3.6 & S1.8 (LLS1)\\
RXS J22287+3335    & 2.6& 22 & 5 & 3 & 2.8 & & & & 4.2 & S1 \\
NPM1G-04.0637      & 2? & 4  &   &   & 3 & & & & 7 & S1 \\

\hline\hline

\end{tabular}
\end{table*}

\subsection{Profile variations}

A few objects have been observed spectroscopically more than once (Table 1). Comparing spectra 
from different epochs could be of particular interest to study the line profile variability 
of the objects. Furthermore, a visually bright, variable quasar can be a good candidate for 
eventual future reverberation-mapping campaigns. Profile changes are shown in Fig. 4 for two 
objects (NPM1G +27.0587 and RXS J20440+2833). The H$\beta$ region was preferred, unless the quality 
of the spectra was not good enough in this region, so H$\alpha$ was used instead. Before 
subtracting, the spectra were scaled in such a way that no sign of the narrow lines to be seen 
in the resulting spectrum ([OIII] for H$\beta$ and [SII] for H$\alpha$). This guaranties 
that variable atmospheric conditions, which may change artificially the continuum, 
will not alter the broad profiles. One sees that, while for RXS J20440+2833 the results are not 
conclusive enough in terms of profile variations, NPM1G +27.0587 clearly shows a change in 
H$\alpha$ flux (by about 50\%), making it quite suitable for reverberation mapping studies in the future.

\begin{figure}
\centering
\resizebox{\hsize}{!}{\includegraphics{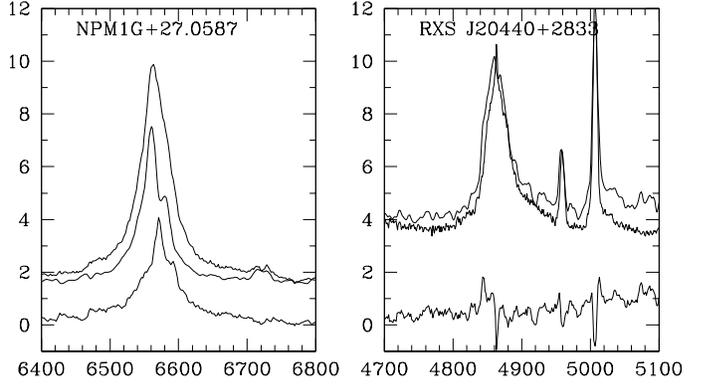}}
\caption{Profile variations for two objects, NPM1G+27.0587 and RXS J20440+2833, H$\alpha$ and H$\beta$ 
regions respectively (see the text). The original spectra taken at two different epochs (Table 1) are 
shown as thick lines and the difference spectrum, as a thin one. Artificial offset may be added 
for a better presentation.}
\label{Figure:4}
\end{figure}

\subsection{Reddening}

Since most of our spectra cover both H$\beta$ and H$\alpha$ regions, we explored the connection 
between the Balmer decrement and the optical continuum slope in more details in order to study 
how the reddening affects the central continuum source and the BLR, especially 
taking into account that many of our objects are heavily reddened. Figure 5 shows the V-band 
absorption affecting the BLR, assuming a cannonical intrinsic broad line decrement of 2.8, 
$A_{\rm V}^{\rm BLR}$, vs. the continuum absorption, $A_{\rm V}^{\rm cont}$, assuming an intrinsic 
optical slope of $a_{\rm \nu}^{\rm intr}=2$, where $F_{\rm \nu} \sim \nu^{a_{\nu}}$. The observed continuum 
slope, $a_{\rm \nu}^{\rm obs}$, was measured between 5200 and 6200 \AA\AA, Table 1,  and then the absorption 
can be roughly estimated as $A_{\rm V} \approx 0.83(a_{\rm \nu}^{\rm intr} - a_{\rm \nu}^{\rm obs})$.
The choice of $a_{\rm \nu}^{\rm intr}$ was made based on the steepest objects (presumably unabsorbed) 
that we have in our sample (see also Dietrich et al.  2005, whose data is also added in Fig. 5), but our main 
findings are not affected by the exact value of  $a_{\rm \nu}^{\rm intr}$ (see also Sect. 4.3). 
As one may also see from Fig. 5, $A_{\rm V}^{\rm cont}$ and $A_{\rm V}^{\rm BLR}$ correlate rather well 
(Pearson correlation coefficient of 0.71), especially taking possible uncertainties in both directions. 
It is interesting to note, however, that the slope of the best fit is not 1 as one can expect, but 
close (within errors) to 2 instead; Sect. 4.3 discusses this issue.

\subsection{Individual notes}
Since most of the objects from our sample have not been previously discussed in the 
literature, here we consider some of their most interesting features.

2MASX J005050+3536.
A spectrum of this object of slightly worse spectral resolution is published by 
Wei et al. (1999). Both spectra look quite alike in terms of broad line profiles, 
continuum level, etc. See also Grazian et al. (2000) for an additional spectrum.

MCG+08.17.060.    
Due to unfavorable atmospheric conditions, the quality of this spectrum is rather 
low and no attempts to subtract iron or host galaxy have been made. For the same 
reason, the H$\beta$ profile has not been measured and no $M_{\rm BH}$ estimated.
The active galaxy is a component of an interacting pair. Kollatschny et al. (2008) have recently 
published a better spectrum of this object. Their measurement of FWHM(H$\alpha$) is similar 
within the errors. The FIRST radio emission ($\approx$18 mJy; Bauer et al.  2000) is not 
sufficient to place the object into the radio-loud category.

RXS J11401+4115.  
This object again has a rather noisy spectrum, so the FeII measurements are rather insecure. 
The SDSS spectrum of this object clearly indicates an S1 galaxy with strong FeII emission, 
but also potentially strong host galaxy continuum. The Strasbourg CDS Simbad database gives 
V=13.8 for this object, but our observations suggest the object should be at least a magnitude 
fainter. Even fainter, the continuum appears to be on the SDSS spectrum. The level 
of detected radio emission (2 mJy) from this object places it well within the radio-quiet 
category.

PG 1211+143.      
This is a well known NLS1 quasar. Reverberation mapping studies (Peterson et al.  2004) infer a central
mass, rather similar to ours (Table 1) even though FWHM(H$\beta$) and the continuum
level are slightly different (see also Kaspi et al.  2005).

RXS J12308+0115.
AThis is a radio-quiet NLS1 object with strong FeII. Located very close to 3C 273 in the sky.
In a recent paper, Landt et al. (2008) report H$\beta$ flux and continuum measurements 
twice as high as ours, a possible indication of significant variability, 
yet the width of the line remains the same within $\sim 4\%$.  

RXS J16312+0955.
This compact RASS source has a low-dispersion spectrum, published in Grazian et al. (2000). 

RXS J17233+3630.
This active galaxy is a part of an interacting pair. De Grijp et al.  1992, published a 
spectrum of the objects of a slightly worse quality. The detected radio-emission of 3 mJy 
(Wadadekar, 2004) classifies the object as radio-quiet. The ratio of [SII]$\lambda6716$ 
and [SII]$\lambda6731$, a doublet separated in this spectrum, is a good density indicator 
and implies $N_{\rm e}\sim1000$ cm$^{-3}$ for the NLR (e.g. Ostebrock, 1974).
Due to the variable atmospheric conditions when the spectrum was taken, all the continuum and 
flux measurements (Tables 2 and 3) should only be considered as a rough estimate.

IRAS 18423+2201.
Moran et al. (1996) published a spectrum of this object, where the broad component 
of H$\beta$ appears much stronger and probably broader. The low level of radio 
emission (2 mJy; Condon et al.  1998) classifies the object as radio quiet. Although 
the broad lines are clearly seen in the spectrum, the place of this object on the 
diagnostic diagrams (e.g. [OIII]/H$\beta$ vs. [OI]/H$\alpha$ and [OIII]/H$\beta$ 
vs. [SII]/H$\alpha$) appears to be on the border between the Seyferts and the LINERs. 

NPM1G+27.0587.
This is an IRAS object behind the Milky Way with a published spectrum around H$\alpha$ 
(Takata et al.  1994).

RXS J20440+2833.  
This is an object behind the Milky Way (Motch et al.  1998)

NPM1G-05.0589.
This spacially extended object (galaxy) does not show broad lines with the exception 
of a very broad wing of H$\alpha$. The host galaxy determines the optical 
continuum entirely. All ratios -- [OIII]/H$\beta$, [OI]/H$\alpha$, [SII]/H$\alpha$, and 
[NII]/H$\alpha$ indicate clearly Seyfert galaxy, so we classified the object as S1.9.
The ratio or the [SII] doublet lines implies an NLR density of about 700 cm$^{-3}$.
A low-dispersion spectrum of the object is also available in Grazian et al. (2002).

RXS J21240-0021.
It is another object, where the host galaxy continuum appears to significantly dominate 
the AGN continuum; however, broad wings (H$\alpha$, H$\beta$, HeII$\lambda 4686$) are 
clearly seen. The place of the object on the diagnostic diagrams is on the border 
between Seyfert-HII-LINER galaxies, suggesting the possible composite nature of its spectrum. 
Most likely the object should be classified as a low luminosity S1 galaxy. The [SII] ratio 
of 1.15 implies an NLR density of about 600 cm$^{-3}$. The SDSS spectrum and a low 
dispersion one in Grazian et al. (2002) are available as well. 

NPM1G+24.0470.  
This spacially extended object probably also has a composite nature, being on the 
border between Seyfert-HII-LINER regions on the diagnostic diagrams. Weak broad lines 
are probably present in the H$\alpha$ profile, which technically means the object should 
be classified as S1.9. The host galaxy is mostly contributing to the continuum.
The [SII] ratio of 0.88 implies a density of about 2000 cm$^{-3}$.

RXS J21592+0952.
This is a broad-line object that appears to have a double-peaked profile (e.g. 
Eracleous \& Halpern, 1994; Strateva et al.  2003), clearly seen in 
H$\beta$. The object was also observed by Wei et al. (1999), whose spectrum reveals 
stronger broad emission. It is a weak radio source, well within the radio-quiet range.

RXS J22027-1304.  
This object is a spatially extended galaxy. On the diagnostic diagrams, it stays 
well inside the Seyfert region and should be classified as either S1.5 or S1.8. 
No signatures of FeII are seen.

RXS J22160+1107.  
Here the continuum is largely dominated by the host galaxy. Diagnostic ratios suggest 
the Seyfert nature of the object; and taking into account the weak broad wings, we 
classify the object as low-luminosity S1 or S1.8.  

RXS J22287+3335.  
This is a rather bright S1 type object with moderate FeII emission. Wei et al. (1999) 
published a spectrum of slightly worse quality, on which the object, interestingly, 
appears to be much fainter ($\approx$5 times in terms of the continuum level), 
indicating possibly strong variability and H$\beta$ appears to be narrower there as well.

NPM1G-04.063.
This is a very interesting object with lines of significant width but continuum -- clearly 
dominated by the host galaxy. If the starlight correction is correct, the intrinsic 
AGN continuum appears to diminish towards the shorter wavelengths, down to zero 
around H$\beta$, which suggests significant absorption between the continuum 
source (e.g. the accretion disk) and the BLR. Some of this absorption may also 
be associated with the BLR itself, due to the quite large Balmer decrement for 
the broad H components ($\approx$6).

\begin{figure}
\centering
\resizebox{\hsize}{!}{\includegraphics[angle=270]{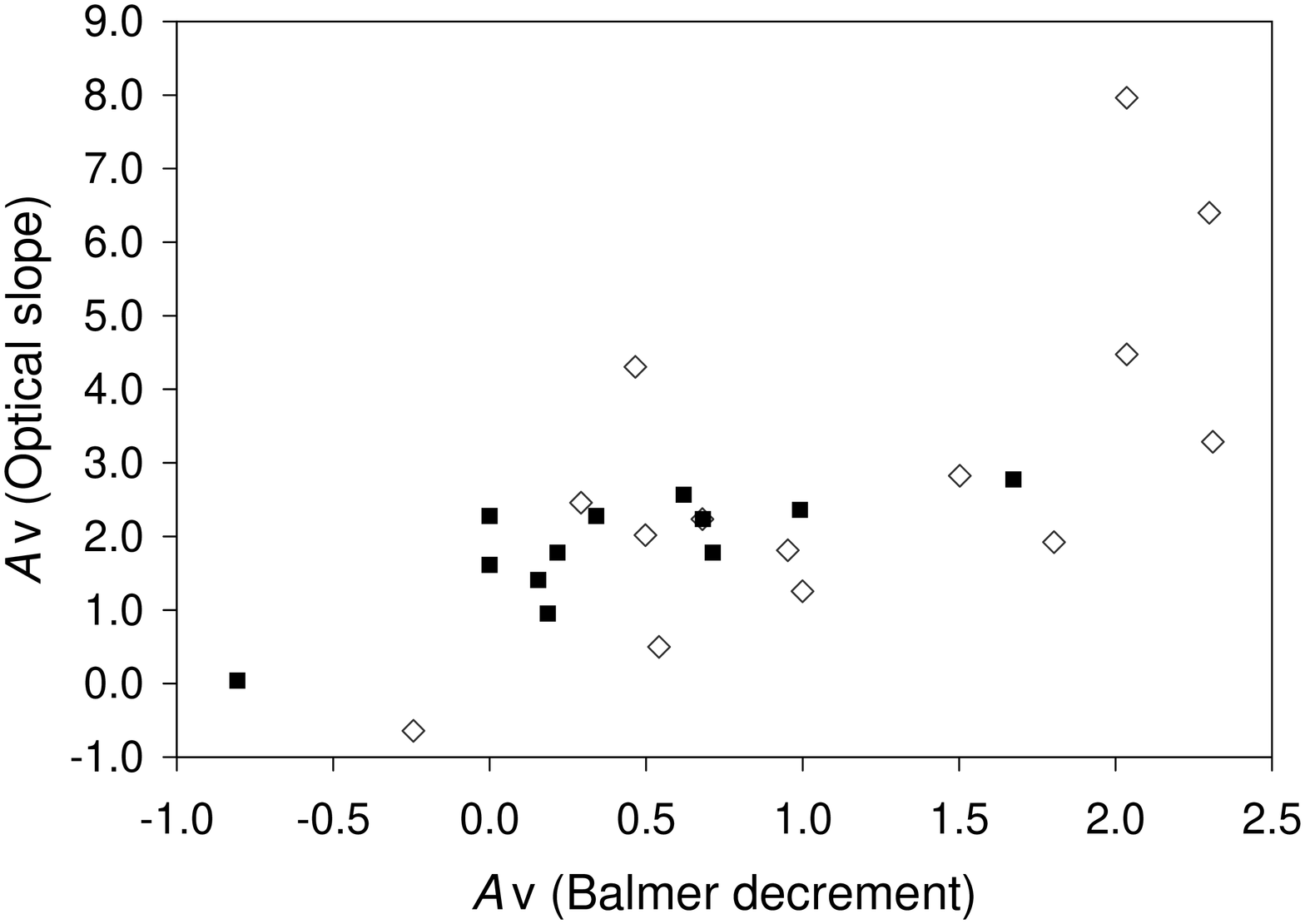}}
\caption{A comparison of the absorption indices ($A_{\rm V}$), calculated through the 
Balmer decrement of the broad components and the optical slope, assuming an unabsorbed 
slope of $a\simeq2$ (see the text). The open symbols represent our data, the filled ones 
are NLS1's from Dietrich et al. (2005). The typical error of $A_{\rm V}$ should be
in order of $\pm 0.5$ in both directions.}
\label{Figure:5}
\end{figure}

\section{Discussion}

\subsection{Line profiles}
While fitting the line profiles, we did not follow a multi-Gaussian/Lorentzian approach, because 
two or three, say Gaussians, fitting the broad profile can hardly have any physical meaning 
(in the sense that they can hardly represent different physical regions). 
Furthermore, each complex profile can be fitted to the desired level of accuracy with a large 
enough number of functions; however, the fit is not necessarily meaningful, not even unique in 
terms of degeneracy. We have to mention, though, that multi-Gaussian fits are 
often used by other authors, mainly for the sake of a better description of the profile 
(Warner et al.  2003; Popovic et al.  2004; see also Bachev et al.  2004, and the discussions there).  
The impression that the narrower profiles are better described by a Lorentzian, not Gaussian,
(Sulentic et al.  2000; V\'{e}ron-Cetty et al.  2001), however,
may have no physical meaning, since it is not clear to what extent FWHM measures physically 
meaningful velocity-related quantity in a complex profile. A more adequate measure would be 
the second moment of the profile (the velocity dispersion), which is infinity for a 
Lorentzian. Even if this were not the case (say for a modified Lorentzian), a Lorentzian-type 
profile typically has higher dispersion than a Gaussian of the same FWHM. Therefore, it would 
be biased to conclude that from two profiles of the same velocity dispersion, the narrower 
one (in terms of FWHM!) resembles a Lorentzian more than a Gaussian.

\subsection{The obscured AGN population}
Finding more and more rather bright (i.e. 14--15 magnitude) AGN shows clearly that 
color-based quasar surveys, like PG for instance, are not complete. This is obviously 
so, because the quasars are very different, not only in terms of emission-line, radio, 
X-ray properties, but also in terms of the optical continuum slopes, which ultimately 
define the colors. The intrinsic quasar continuum obviously may not only depend on the 
exact accretion solution type and accretion parameters, such as black hole mass, accretion 
rate, accretion disk inner and outer radii, etc. The continuum shape and level can be 
severely altered by the host galaxy contribution and the internal absorption, and
we found evidence of both in the sample we studied. The 
reddened continuum slopes can eventually explain why such bright object are not 
discovered earlier as AGN. On the other hand, a correct assessment of the 
intrinsic $\lambda L_{5100}$ is tightly linked to the estimates of the bolometric 
luminosity and the central mass through empirical relations (the often used 
$L_{\rm bol}\simeq 9 \lambda L_{5100}$, see also Sect. 3.2). 
It also affects other characteristics, like optical-to-X-ray slopes, radio-loudness (e.g. 
Ho, 2002), line equivalent widths, etc. Therefore, one needs to know the 
intrinsic quasar continuum for accurately computing the accretion parameters and 
exploring correlations like EV1. For instance, large quasar samples with well-studied 
spectral properties (e.g. Boroson \& Green, 1992; Marziani et al.  2003) rarely make 
an attempt to assess the real continuum level through correcting for the starlight 
and the intrinsic absorption, presuming that their contribution is small. It may indeed 
be so for some bright quasars, but as we see here (see also Dietrich et al, 2005), this 
is not always the case. Furthermore, dangerous biases may be introduced -- the host 
galaxy will contribute to the continuum mostly for a lower-luminosity AGN, and the internal 
absorption may be connected with the orientation if associated with a thick torus, 
and the broad-line widths if the torus is coplanar to a flattened BLR.   
It should be easy to correct for the host galaxy in a spectrum of relatively good 
quality. Unfortunately, the host galaxy contribution depends largely on the slit 
size, seeing, etc. and therefore may differ significantly from observation to 
observation and from instrument to instrument. This also can have consequences 
for the reverberation-mapping studies, in the sense that a variable starlight 
contribution can at least add some extra noise in the cross-correlation function 
between the continuum and the lines.

\subsection{Where is dust located?}

In this paper we did not make any attempt to correct the continuum for the intrinsic 
absorption; nevertheless, the latter is rather apparent for objects like 
NPM1G-04.063, where quasar continuum around H$\beta$ appears to be totally absorbed. 

One may make an attempt to take this absorption into account either by measuring the 
continuum slope or by comparing the Balmer decrement with the theoretical expectations 
(i.e. case ``B'' recombination). Both approaches may lead to incorrect results, however. 
The slope can, of course, vary for a number of reasons and the dust reddening is only 
one of them. On the other hand, the Balmer decrement can also be misleading since we do 
not know exactly where the absorbing material is. If it is uniformly spread outside 
the NLR, then such a correction can in principle be performed, and the absorption will 
presumably equally affect the NLR, the BLR, and the central continuum source. 
Different Balmer decrements for the broad and the narrow lines, as seen in some of 
the objects in our sample, suggest this may not always be the case. If the absorbing 
material is at least outside the BLR, then the broad line decrement could be 
used to assess the unabsorbed continuum level and shape. Unfortunately, we cannot 
rule out the presence of absorption inside the BLR inner radius (or mixed with 
the BLR matter), in which case the continuum will be affected, but indications 
for this might not be seen from the emission line ratios.

An important clue to solving the problem of the dust location can be provided by the relation 
between the optical continuum slope and the Balmer decrement. We found a correlation between 
the slope and the decrement absorption indices (Fig. 5). One is to note, however, that the wide 
range of both -- Balmer decrements and continuum slopes observed in our rather small sample 
is not typical. For instance, Dong et al. (2008) report a much narrower range of Balmer decrements 
for a large sample of SDSS quasars -- typically between $-0.5$ and 1, 
and slopes -- between 1.2 and 2.2 (both expressed in terms of $A_{\rm V}$). Their sample is, 
however, blue-selected and is obviously biased against the obscured AGN population. 
This may explain why Dong et al. (2008) find no correlation between the decrements and the slopes in their sample.

If further confirmed, the relation  $A_{\rm V}^{\rm cont}\simeq 2 A_{\rm V}^{\rm BLR}$ (Sect. 3.4) 
suggests that the reddening on average affects the center more than the BLR. Even adopting a different 
value for the intrinsic continuum slope (like the often assumed $a_{\rm \nu}^{\rm intr}\simeq 1$) only 
offsets the relation between the Balmer decrement and the continuum slope, but does not change the slope 
of the relation. Adopting a different value for the intrinsic Balmer decrement (e.g. 2.8 instead of 3.1) has a similar effect. 
One way to explain this result is to assume that the dust is associated with the BLR matter and the 
reddening affects the continuum coming from inside more than the BLR emission itself. Another 
explanation would be that the dust is much farther outside the BLR (probably associated with a torus), 
and we see the active nucleus just above the torus edge. 
As a result, the central continuum can appear heavily absorbed, but the BLR light is not that 
affected, taking their different spatial dimensions into account. 
If the latter explanation turns out to be the correct one, studying the line profiles of 
such partially obscured objects can be important for probing the exact BLR geometry and kinematics, 
since knowing the torus opening angle, the inclination can be estimated for these objects.



\section{Summary}
We obtained optical spectra of bright AGN, mostly recently discovered 
as X-ray sources. Different emission line parameters were measured, including 
important EV1 characteristics, such as FWHM(H$\beta$) and $R_{\rm Fe}$. Black hole 
masses were estimated through an empirical relation, invoking the line width 
and the continuum level.
We show that the host galaxy contribution to the continuum may be non-negligible 
in many cases, often about 50\% or more. The host galaxy, as well as the 
intrinsic absorption, can significantly alter the continuum slope, offering a 
possible explanation why such bright object have not been discovered earlier as AGN.
Based on the line widths and emission line ratios, we provided a Seyfert type 
classification of each object, showing that, although previously listed as S1 
types, some objects should be considered at least S1.9, if not S2 or LINERs.  
Finally, taking into account that many of our objects are heavily reddened, 
we discuss different possibilities about the exact location of the absorbing material. 

\begin{acknowledgements}
The authors are grateful to Paola Marziani for carefully reading an earlier version of the manuscript 
and for her valuable comments. We thank Prof.~Y.~Papamastorakis, Director of the Skinakas Observatory, 
and Dr.~I.~Papadakis for the allocated telescope time. The design and manufacturing of the the Focal 
Reducer Rozhen (FoReRo) were performed in the workshop of the Institute of Astronomy, Bulgarian
Academy of Sciences, with financial support by the Ministry of Education
and Science, Bulgaria (contract F-482/2201).
This research made use of the CDS-Strasbourg (Simbad), NASA-IPAC (NED) databases, 
and of the IRAF package distributed by the NOAO.

\end{acknowledgements}

\end{document}